\documentclass[acmtois]{acmtrans2m}
\usepackage[final]{epsfig}\usepackage{amsmath}

\newcommand{\avg}[1]{\left\langle{#1}\right\rangle}

\newcommand{\be}{\begin{equation}}
\newcommand{\ee}{\end{equation}}
\newcommand{\beas}{\begin{eqnarray*}}
\newcommand{\eeas}{\end{eqnarray*}}
\newcommand{\bea}{\begin{eqnarray}}
\newcommand{\eea}{\end{eqnarray}}
\newcommand{\req}[1]{(\ref{#1})}
\newcommand{\comment}[1]{}
 
\title{When are recommender systems useful?}
\author{MARCEL BLATTNER \\ ALEXANDER HUNZIKER  \and PAOLO LAURETI 
 \\ Department of Physics, University of Fribourg}
\begin{abstract}
Recommender systems are crucial tools to overcome the information
overload brought about by the Internet. Rigorous tests are needed to
establish to what extent sophisticated methods can improve the quality
of the predictions. 
Here we analyse a refined correlation-based collaborative filtering algorithm
and compare it with a novel spectral method for recommending. We test them on
two databases that bear different statistical properties
(MovieLens and Jester) without filtering out the less active users and
ordering the opinions in time, whenever possible.
We find that, when the distribution of user-user correlations is narrow,
simple averages work nearly as well as advanced
methods. Recommender systems can, on the other hand, exploit a great deal
of additional information in systems where external influence is negligible and
peoples' tastes emerge entirely. These findings are
validated by simulations with artificially generated data.

\end{abstract}



\begin{document}

\maketitle

\section{Introduction}
One of the most amazing trends of today's globalized economy is
{\em  peer production}~\cite{An06}. 
An unprecedented mass of unpaid workers is
contributing to the growth of the World Wide Web: some build
entire pages, some only drop casual comments, having no other
reward than {\em reputation}~\cite{MaZh04}.
Many successful websites (e.g. Blogger and MySpace) are just platforms
holding user-generated content.
The information thus conveyed is particularly valuable because it
contains personal opinions, with no specific corporate interest. It is,
at the same time, very hard to go through it and judge its degree of 
reliability. If you want to use it, you need to filter this
information, select what is relevant and aggregate it; you need to
reduce the information overload~\cite{Ma94}.

As a matter of fact, opinion filtering has become rather common on the
web. There exist search engines (e.g. Googlenews) that are able to extract news
from journals, websites (e.g. Digg) that harvest them from blogs, platforms
(e.g. Epinions) that collect and aggregate votes on products. The basic
version of these systems ranks the objects once for all, assuming they
have an intrinsic value, independent of the
personal taste of the demander~\cite{LaMo06}. They lack
{\em personalisation}~\cite{Ke06}, which constitutes the
new frontier of online services.

Users need only browse the web in order to leave recorded traces, the eventual
comments they drop add on to it. The more information you release, the
better the service you receive. 
Personal information can, in fact, be
exploited by {\em recommender systems}.  
The deal becomes, at the same time, beneficial to the
community, as every piece of information can potentially improve the
filtering procedures. Amazon.com, for instance, uses one's purchase
history to provide individual suggestions. If you have bought a
physics book, Amazon recommends you other physics books: this is
called  item-based recommendation~\cite{BrHe98,SaKa01}. 
Those who have experience with it
know that this system works fairly well, but it is conservative as it
rarely dares suggesting books regarding subjects you have never
explored. We believe a good recommender system should sometimes help
uncovering people's hidden wants~\cite{MaZh01}.

{\em Collaborative filtering} is currently the most successful
implementation of recommendation systems. It essentially consists in
recommending you items that users, whose tastes are similar to yours,
have liked. In order to do that, one needs collecting taste
information from many users and define a measure of similarity.
The easiest and most common ways to do it is to use either
correlations or Euclidean distances.

Here we test a correlation-based algorithm and a
spectral method to make predictions. We describe these two families of
recommender systems in section~\ref{met}, and propose some
improvements to currently used algorithms. In section~\ref{exp} we
present the results of our predicting methods on the MovieLens and
Jester data sets, as well as on artificial data. We argue that the
distribution of correlations in the system is the key ingredient to
state whether or not sophisticated recommendations outperform simple
averages. Finally, we draw some conclusions in section~\ref{con}.

\section{Methods}\label{met}
Our aim is here to test two methods for recommending, spectral and
correlation-based, on different data sets.
The starting point is data collection. One typically has a system of
$N$ users, $M$ items and $n$ evaluations. 
Opinions, books, restaurants or any other object can be treated,
although we shall examine in detail two fundamentally different
examples: movies and jokes.
Each user $i$ evaluates a
pool of $n_i$ items and each item $\alpha$ receives $n_{\alpha}$
evaluations, with $n=\sum_{i=1}^N n_i=\sum_{\alpha=1}^M n_{\alpha}$. 
The votes $v_{i \alpha}$ can be gathered in a matrix $V$.
If a user $j$ has not voted on item $\beta$, the corresponding
matrix element takes a constant value $ v_{j \beta} =$ EMPTY, usually set
to zero. 

Once the data collected into the voting matrix, we aim to predicting
votes before they are expressed. That is, we would like to predict if
agent $j$ would appreciate the movie, book or food $\beta$,
before she actually watched, read or ate it.
Say, we predict that user
$j$ would give a very high vote to item $\beta$ if she were exposed to
it; we can then recommend $\beta$ to $j$ and verify {\it a posteriori}
her appreciation. Ideally, we would like to have a prediction for
every EMPTY element of $V$. 

Most websites only allow votes to be chosen from a finite set of
values. In order to take into account the fact that each person adopts an
individual scale, we compute each user's $i$ average expressed vote
$\avg{v_i}$ and subtract it from non empty $v_{i \alpha}$'s. 
\comment{
In the
following we shall consider rescaled votes and denote them with the
same symbols. We shall indicate with $\vec{v}_i$ the vector containing
the votes of
user $i$, which is the $i^{th}$ row of matrix $V$.
}
The methods we analyze give predictions in the following form~\cite{De99}:
\be\label{pred}
v_{j \beta}' = \avg{v_j} + \sum_{i=1}^N S_{j i} (v_{i \beta}-\avg{v_i}),
\ee
where $v_{j \beta}'$ is the predicted vote and $S$ is a
similarity matrix. 
%It is sometimes required that $S_{j i}>0$ and
%$\sum_i S_{j i} = 1$, in which case they are just normalised weights. 
The choice of $S$ is the
crucial issue of collaborative filtering. One has, in fact, very often
to face a lack of data, which makes it difficult to estimate the
similarity between non overlapping users. We shall describe, in the
following, correlation-based and spectral techniques to cope with this
problem.

\subsection{Correlation}
Correlation-based methods for recommending make use of 
user-user linear correlations as similarity measures. If we call
$\avg{v_i}$ the average vote expressed by user $i$,
the correlations $C_{i j}$ can be calculated with Pearson's formula~\cite{NumRec}:
\be\label{pearcorr}
C_{i j} = {{\sum_{\alpha}  (v_{i \alpha}-\avg{v_i}) (v_{j
    \alpha}-\avg{v_j}) } \over {\sqrt{\sum_{\alpha}  (v_{i
      \alpha}-\avg{v_i})^2} \sqrt{\sum_{\alpha}  (v_{j
      \alpha}-\avg{v_j})^2}}},
\ee
with $C_{i j}=0$ 
if users $i$ and $j$ haven't judged more than one item in common.
\comment{
Note that this expression is redundant in our case, as $\avg{v_i}=0$
for all $i$.
}
Unexpressed votes can then be forecast by setting $S_{i j}\propto C_{i j}$ in
eq.~\req{pred}. This estimation is often used as
a rule of thumb, without knowing if it is justified and why. 
There is, at our knowledge, only one model~\cite{BaBe03}
where the convergence of $v_{j \beta}'$ to the real vote, in the limit
of an infinite system, is guaranteed. 

The use of pair correlations alone is often not very effective in
predicting tastes. In fact,
$C_{i j}$ is a measure of similarity between the behavior of two users who
have expressed votes on a number $n_{i j}$ of commonly evaluated items. When
the matrix $V$ is very sparse and $n_{i j}$, as a consequence, small
or zero for many couples of users, such a measure becomes poorly informative.
A popular solution to this problem~\cite{BlZh07}
is to estimate unknown votes via a linear combination
of $v_{j \beta}'$ and a global average, i.e.
$
v_{j \beta}'' \propto q v_{j \beta}' + (1-q) m(\beta),
$
where $q$ is a constant weight between $0$ and $1$ and $m(\beta)$ the
average vote expressed on item $\beta$ by all users. The typical
choice, $q=1/2$, amounts to defining $S_{i j}= 1/n_{\beta} + C_{i
  j}$  in eq.~\req{pred}.

After testing many different versions of
correlation-based recommendations, with the same number of free
parameters, we chose a more effective ansatz: every time $C_{i j}=0$ we replace it
with the average value of the correlation across the population. Such a
mean-field solution improves the results and eliminates,
at the same time, the parameter $q$.
As for the normalization
of the weighted sum in eq.~(\ref{pred}), we found that the best choice is $S_{j
  i}= C_{j i} / \sum_i |C_{j i}| $. Small adjustments can be  made on the
minimal number of common items $n_c$ required to compute
correlations. A low value of $n_c$ may, in fact, improve the average
prediction quality at the price of very large fluctuations. We set
$n_c=3$ in most cases.

Let us stress here that the method we used 
can be further improved by calibrating additional
parameters. 
Case amplification~\cite{BrHe98}, for instance, consists in taking a
similarity measure proportional to some power $\gamma$ of the correlation,
in order to punish low correlation values. Upon optimizing the value
of $\gamma$, the prediction power of our correlation-based method is able
to compete with that of spectral techniques~\cite{Mo07}. 

\subsection{Spectral}
An alternative approach to recommending makes use of spectral techniques~\cite{SaKa00,BiPa98}.
Users can be represented by their vectors of votes $\vec{v}_i$
in a $M$-dimensional metric space $\mathcal{M}$. One can define, between all couples
of users $(i,j)$, an overlap $\Omega_{i j}$ as
a decreasing function of their distance. Users can thus be
represented as nodes of a weighted graph.
Spectral methods have recently been used to detect clusters in
networks~\cite{SeRi03,DoMu04,CaSe05} and can equally be applied to devise
recommender systems. In order to implement them for this purpose, 
we take the following fundamental steps:
{\sl i)} Calculate the overlap matrix $\Omega$ that defines our
weighted graph. 
{\sl ii)} Find the spectrum of its Laplacian after dimensionality reduction.
{\sl iii)} Calculate user similarities using eigenvectors'
  elements, and make predictions with eq.~\req{pred}.
The method is not trivial and each one of the preceding points
contains subtleties. After testing many different options, 
we have used the one that yields the best and more
stable results. 

The definitions and the procedures
we shall describe here, stand on the following assumption:  
if we disposed of the votes of all users on all items, the points of
coordinates $\vec{v}_i$ would
only occupy a compact subspace of $\mathcal{M}$, a manifold-like structure
of dimension $k<<M$. This approach, whose
validity will be verified {\it a posteriori}, is inspired by
ref.~\cite{BeNib}, where the interested reader can find the details.

A preliminary step is the substitution of the EMPTY
entries of the voting matrix with the corresponding object's average
received vote $m(\alpha)$. 
We then define the elements of the overlap matrix as
$
\Omega_{ij} = \exp(- d_{i j}^{2}/\Gamma^2)
$,
where  $d_{i j}=||\vec{v}_i-\vec{v}_j||$ is the Euclidian distance between the
vectors of votes of users $i$
and $j$.  
$\Omega$ gives higher weights to pairs of users whose votes are
closer, and reaches its maximum $\Omega_{ij}=1$ when the distance is $0$.
The external parameter $\Gamma^2$ controls the size of the neighborhood,
since $\Omega_{ij} \to 0$ for $d_{i j}\gg \Gamma$. The
performance of our experiments was almost unchanged within a wide
range of this parameter. We fixed $\Gamma=\max_{i,j} d_{i j}$, which
allows to develop the exponential $\exp(-
d^{2}/\Gamma^{2})\simeq 1-(d/\Gamma)^2$. Hence we set
%\be\label{omega}
$$
\Omega_{ij} = 1 - \left(\frac{||\vec{v}_{i} -
  \vec{v}_{j}||}{\max_{i,j}||\vec{v}_{i} -
\vec{v}_{j}||}\right)^2.
$$
%\ee

\begin{figure}%[ht!]
\centerline{\includegraphics[width=0.7\textwidth]{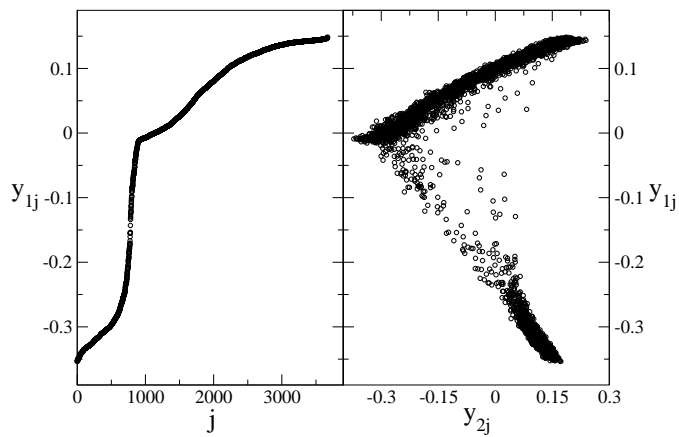}}
\caption{Identification of clusters of users in Jester with spectral
  analysis. {\sl Left graph:} $10\times$value of the elements of the
  first non-trivial eigenvector $\vec{y}_1$ of the normalized
  Laplacian matrix, plotted against their cardinality. The clear
  discontinuity around $0$ indicates the presence of two big
  clusters. {\sl Right graph:} The first non-trivial eigenvector is
  plotted here against the second. The presence of the two main
  clusters is clearly visible.
}
\label{jeseig}
\end{figure}

In order to obtain the optimal embedding in $k$ dimensions of the
structure formed by users in the space of the votes, 
ref.~\cite{BeNib} prescribes to find
%$E = \sum_{ij} \Omega_{ij} ||v_{i} - v_{j}||^{2}$~\cite{belkin2}.
the first $k$ eigenvectors of the normalized
Laplacian $L=D^{-1/2}(D -\Omega)D^{-1/2}$ of $\Omega$, 
where $D$ is its diagonal weight matrix, of elements
$D_{ii}= \sum_{j}\Omega_{ij}$.
When used for partitioning the population of users, the spectrum of
the normalized Laplacian $L$ has revealed more effective and
stable than that of other matrices~\cite{LuBe04}.
Let $\vec{y}_{0},.....,\vec{y}_{k-1}$ be its first $k$ eigenvectors,
in order of increasing eigenvalues. Note that $\vec{y}_{0}$
always has eigenvalue $0$. The other non-trivial eigenvectors contain
information about eventual subgraphs~\cite{SeRi03}. 
If the graph is connectected and
bipartite, for instance, the components of
$\vec{y}_{1}$ will be positive for one subgraph and negative for the
other. Whenever the two subgraphs are not very well separated, the
distinction between the two components becomes progressively fuzzier.  
An example is given by the left graph of fig.~\ref{jeseig}, where the
components $y_1(j), j=1,2,...,N$ 
of the first non-trivial eigenvector of the Laplacian of
the Jester database show a clear discontinuity. Higher eigenvectors
contribute to define the first two clusters and can reveal the
contours of more weakly connected blocks~\cite{SiEr04}. The right graph of
fig.~\ref{jeseig}, in fact, shows the projection of the Jester matrix
on the first two non-trivial eigenvectors. The presence of two islands
can be detected by eye inspection, confirming that Jester users can be
grouped in, at least, two different categories. 

Let us now come back to the application of our spectral analysis to
recommending systems. Each user $i$ can be represented by a point in
the $(k-1)$-dimensional subspace made of the $i$th components of the first
$k$ eigenvectors, i.e. $\vec{y}(i) =({y}_{1}(i),...,{y}_{k-1}(i))$.
One can think that each one of these coordinates contains information
about the degree of participation of user $i$ in a subgroup of users.
The parameter $k$, which plays the key role in dimensionality
reduction, has to be determined by a cross checking procedure:
we measure the performance of the algorithm on the training set for
different values of $k$, and choose the one which supplies the best
results. In our experiments this is of order $10$.

Finally, we calculate the similarity $S_{ij}$ for each pair of users. 
After comparing many different measures,
the cosine turned out to work best. Thus we define
$$
S_{ij} = \frac{ \vec{y}(i)^T \vec{y}(j)}{||\vec{y}(i)|| ||\vec{y}(j)||},
$$
where the superscript $T$ denotes the transposed of a vector, and
$||\cdot||$ is the ordinary  $L_{2}$ norm.
Here $S_{ij}$ can be
interpreted as an overlap between the participation ratio of two users
to different groups of taste.
Armed with this similarity matrix, we predict votes according to eq.~(\ref{pred}).
Our spectral technique, although very tedious, performs better than
the other methods we tested. 

\section{Experiments}\label{exp}
The purpose of this paper is to evaluate different collaborative
filtering algorithms, and to establish when they can be used
effectively. To this end, we have tested the methods described above
on two data sets, carrying completely different features: MovieLens and
Jester. In order to achieve a better understanding of the role played by
correlations in the votes, we have also made simulations on artificially
generated data. Prior to presenting our results, we shall describe
the data sets used for the experiment.

\subsection{Data sets}
{\bf MovieLens} (movielens.umn.edu) %\cite{MoWeb} 
is a webservice of the GroupLens project (grouplens.org),%\cite{GrWeb}, 
that recommends
movies. Users ratings are recorded on a five
stars scale and contain additional information, such as the time at which an evaluation has been made. 
The data set we downloaded contains $6040$ users $\times$ $3952$
movies, where only a fraction $\eta_M=0.041$ of all possible votes has actually been expressed.
{\bf Jester} (shadow.ieor.berkeley.edu/humor)%\cite{JeWeb} 
is an online joke recommendation
system. It contains $73421$ users $\times$ $100$ jokes and a fraction
$\eta_J=0.55$ of expressed votes. Users ratings are real numbers ranging from $-10$ to $10$. 

While both websites collect ratings of users on items, they 
differ substantially in many respects: the range of allowed ratings $R$,
the users-to-items ratio $N/M$, the sparsity of the voting
matrix and the distribution of votes, among others. In fact,
while the MovieLens data set is roughly symmetric,
the Jester one is heavily asymmetric, with users outnumbering items by
a factor $734$. This is because of the
fixed, low number of jokes ($M_J=100$) one can evaluate in Jester. For the
same reason, the MoviLens data set is much sparser than Jester. In
fact, defining the sparsity coefficient as $\eta=n/(M\times N)$, where
$n$ is the number of recorded ratings, one has
$\eta_M\simeq 4\%$ and $\eta_J\simeq 55\%$. Note that $\eta$ decreases as the
matrix gets sparser, and not {\it vice versa}.

\begin{figure}%[ht!]
\centering
\includegraphics[width=0.7\textwidth]{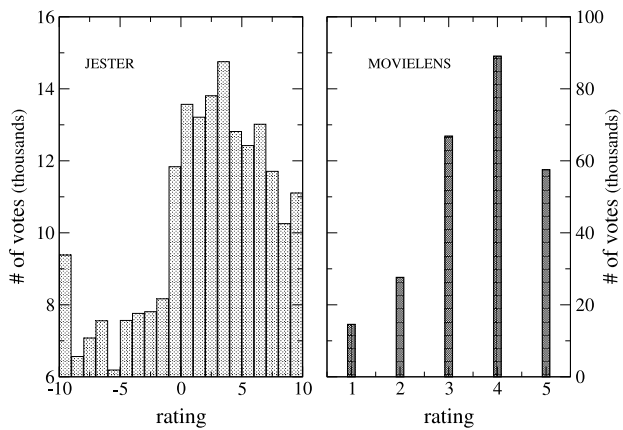}
\caption{Distribution of votes for Movielens and Jester.}
\label{fig0}
\end{figure}
The most fundamental difference, though, is the
amount of {\it a priori} information provided to users. People choose the
movies they want to watch on the basis of a  preliminary
selection. They know actors and directors, read reviews and are
exposed to advertisements. When they actually buy the entrance ticket,
they have some motivated expectactions.
Accordingly, in the MovieLens data set, the distribution of all the issued
votes is unimodal and noticeably shifted towards the positive region (see fig.~\ref{fig0}, right panel). 
No preselection, on the other hand, is possible with online jokes, giving
rise to a more uniform distribution of votes.
This can be verified by arranging the votes on a $5$
beans histogram and computing the entropy $S=-\sum_i p_i \log_5
p_i$, which yields
 $S_J=0.98$ for Jester and $S_M=0.90$ for MovieLens.  
In addition to that, the distribution of votes is slightly bimodal
in Jester, as shown in the left panel of fig.~\ref{fig0}. This suggests the existence of groups of
users with similar taste, which is confirmed by fig.~\ref{jeseig}, as
already pointed out.
In order to gain more insight into this fundamental difference, we also
report, in fig.~\ref{corr_dist}, the distribution of user-user
correlations, the effects of which will be discussed in section~\ref{simu}.

The size of the data sets has been reduced by roughly $50$\% in both dimensions. 
As the cancellations have been done randomly, the statistical properties
of the original data have been preserved. In particular, we tried to maintain the probability
distribution of the number of votes per users, as well as the sparsity
and the $N/M$ ratio. We want to stress that this is crucial when
testing the performance of predictive algorithms on real data in an
objective way. In fact, many experiments can be found in the
literature that only test recommender systems on dense voting matrices.
Typically, users who have judged too few items are struck out, as well
as items that have received too few votes. 
We did not comply to such an habit and made
an effort to {\em keep the filtering level as low as
    possible}, although
this makes predictions much more difficult. 

Once filtered, the data are divided into a training
and a test set. The training set is composed of the data one actually uses
to make predictions on the missing evaluations contained in the test
set. This last is only employed afterwards, to compare predictions and realised
evaluations. We have chosen test sets of dimension $n_{test}=10^4$ for
both databases. The experiments have been carried out as
follows. First, we fix the test set and never change it through the
simulation. Then we progressively fill the training set over time and
make predictions on the entire test set at fixed time steps.

Many different accuracy metrics have been proposed to assess
the quality of recommendations (see ref.~\cite{HeKo04}), one of the most common
of which being the Mean Average Error:
\be\label{mae}
MAE=\frac{1}{n_{test} R}\sum_{j,\beta} |v_{j \beta}'-v_{j \beta}|,
\ee
where the sum runs over all expressed votes in the test set and
$R=|v_{max}-v_{min}|$ is the size of the domain of all possible ratings.
In our experiments, the MAE is calculated, at different sparsity
values $\eta$, on a unique test set. The results for our sets of data
will be presented in the following sections.
\begin{figure}%[ht!]
\centerline{\includegraphics[width=0.7\textwidth]{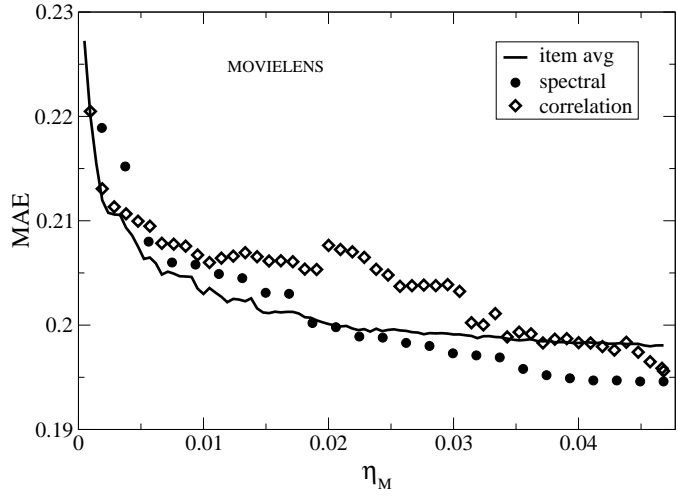}}
\caption{MAE of spectral (with $k=20$), 
correlation-based methods and movie average $m(\alpha)$,
  as a function of $\eta_M$, for the Movielens
  reduced data set ${V}_M$. The sparsity coefficient $\eta_M$ is increased
  according to the timestamps of the expressed votes. The test set is
  composed by the last $10^4$ evaluations.}
\label{fig1}
\end{figure}

\subsection{MovieLens}

After the filtering procedure, we cast the data set
in a voting matrix $V_M(N \times M)$, with $N=3020$ and $M=1976$.
As previously mentioned, the MovieLens database contains the time at
which evaluations have been made. {\em We have sorted the
votes  according to their relative timestamp}, both in the training
set ${V}_M$ and in the test set, which is composed of the last $10^4$
expressed votes. Such a choice is intended to
reproduce real application tasks, where one aims to predict future
votes --which is, of course, much harder than predicting randomly picked
evaluations. It is somewhat less realistic to fix the test set once
for all, but this has the advantage to allow for more objective
comparisons of the results.

The training set has been filled, as well, by adding one vote at a
time, according to the temporal ordering. Predictions have been made at
fixed sparsity values, as shown in fig.~\ref{fig1}. The MAE
obtained with spectral and correlation-based methods are compared
therein. The solid line is the MAE of
predictions made by taking the average vote received by each
movie $m(\alpha)$. Surprisingly, the results achieved with this naive
estimator  are
comparable to those of the sophisticated methods, and outperform them
in the very sparse region. Note that the movie average predicts the
same vote for
every user, while the other methods produce personalized
recommendations. Their utility only emerges after a crossover value $\eta_M\simeq 0.5$.
For most fillings, our spectral method (diamonds) performs
slightly better than our correlation method (circles), which also
suffers of stronger fluctuations.

\subsection{Jester}

\begin{figure}%[ht!]
\centerline{\includegraphics[width=0.7\textwidth]{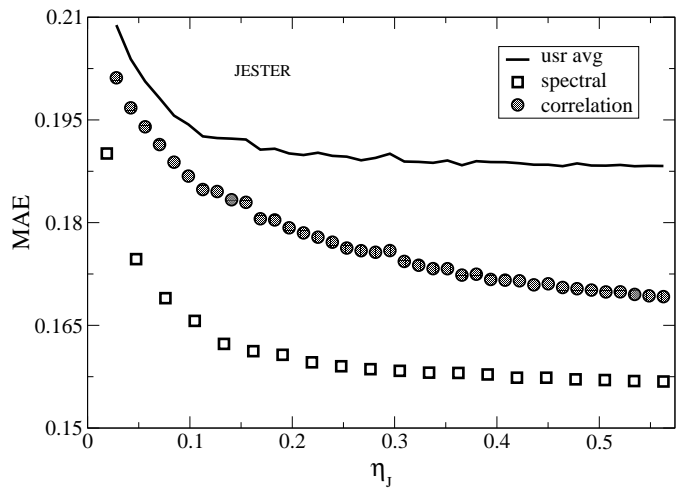}}
\caption{MAE of spectral (with $k=8$), 
correlation-based methods and user average $\avg{v_i}$,
  as a function of the sparsity, for the Jester data set ${V}_J$.
Here $\eta_J$ is increased by random addition of votes. The test
set contains $10^4$ randomly picked evaluations.
}
\label{fig2}
\end{figure}

After reduction, we are left with a data set that can be cast
in a voting matrix $V_J(N \times M)$, with $N=3671$ and $M=100$.
The test set has been fixed once for all by random choice of $10^4$
evaluations. The training set has been filled randomly and predictions
have been produced at increasing $\eta$ values. The results are shown
in fig.~\ref{fig2}. While in the MovieLens data set the item's average
received vote $m(\alpha)$ is a good predictor for all users, here it is not at all
the case. A much better estimate can be produced by
the average vote a user has given to any item, represented
by the straight line in fig.~\ref{fig2}. In fact, in this case,
votes are given on an individual basis: both the absolute opinions and the rating
scales differ severely from user to user. This might partly explain the fact that
sophisticated methods enjoy an edge over simple
averages. Our spectral method (squares) performs much better than our
correlation method (circles), which, in turn, beats the user average
$\avg{v_i}$ by a large amount. 

All methods give rise to a smaller error in the Jester than in the
MovieLens set. This is due to various factors. First, $V_M$ is sparser
than $V_J$. Comparing the predictions at the same level of sparsity,
though, Jester remains more predictable (with our spectral method), in spite
of the much smaller size of its data set. A second factor is
represented by the choice of the test set, which is not random in the
MovieLens case. Finally, the average correlation of Jester users is
higher than that of MovieLens. A more detailed explanation needs an
analysis of the entire distribution of correlations, which is the
object of the next section.

\subsection{Simulations}\label{simu}
The performance of any recommender system depends on the structure of
the data set under investigation. Upon assuming that the user-user correlation is
the relevant variable, the shape of its distribution can give us
some preliminary information. Since correlations are a measure
of similarity of people's tastes, it is trivial to understand that, 
when all users are
equally correlated, the item's average received vote is the best
predictor. When the distribution of correlations is broad, on
the other hand, it becomes useful to make individual predictions,
giving more weight to highly correlated mates. 
For simplicity, the analysis can be
restricted to the mean and the standard deviation of the distribution
of correlations. A higher absolute mean increases the predictability;
a broader distribution enriches the information encoded and requires
clever methods to be exploited.

\begin{figure}%[ht!]
\centerline{\includegraphics[width=0.7\textwidth]{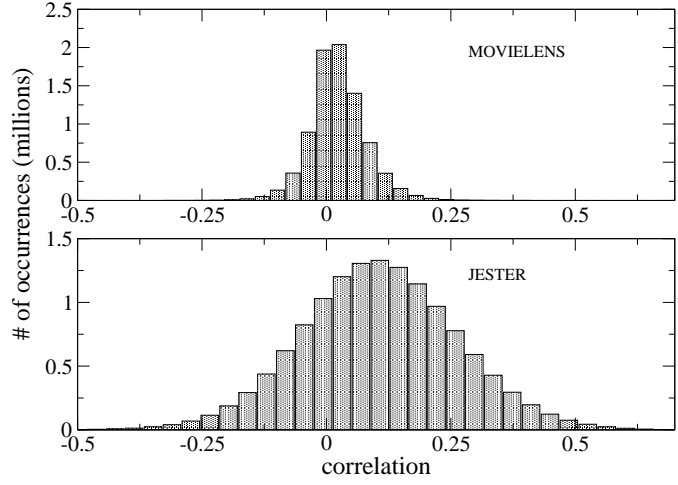}}
\caption{Distribution of correlations for Movielens and Jester.}
\label{corr_dist}
\end{figure}

As a preliminary check of our analysis, we can look at the
user-user correlations, as calculated
from eq.~(\ref{pearcorr}), of our two data sets. It is evident from
fig.~\ref{corr_dist} that Jester has a higher mean
correlation ($\mu_J\simeq 0.1$) than MovieLens ($\mu_M\simeq 0.02$), in accordance with
the fact that Jester allows for better predictions. It also appears that
Movielens correlations have a lower standard deviation
($\sigma_J\simeq 0.16$
vs. $\sigma_M\simeq 0.05$). This explains, in our framework, why sophysticated
methods give much better results than simple averages in Jester
(see fig.~\ref{fig2}) and not in the case of MovieLens (see
fig.~\ref{fig1}). 

To test our hypothesis in a systematic way, we generate artificial votes, where we control the
structure of the correlation distribution, according to the following procedure.
First, we create a valid correlation matrix $\bar{C}$ of fixed size $N \times N$,
with the desired mean $\mu$ and variance $\sigma^2$, as explained in
appendix~\ref{app1}.
Then we draw a multivariate Gaussian distribution of votes $\bar{V}(N \times
M)$, with $\bar{C}$ as input correlation 
matrix. Finally, we perform predictions on these artificially generated
data, comparing our correlation-based method with simple averages.
\begin{figure}[ht!]
\centerline{\includegraphics[width=0.6\textwidth]{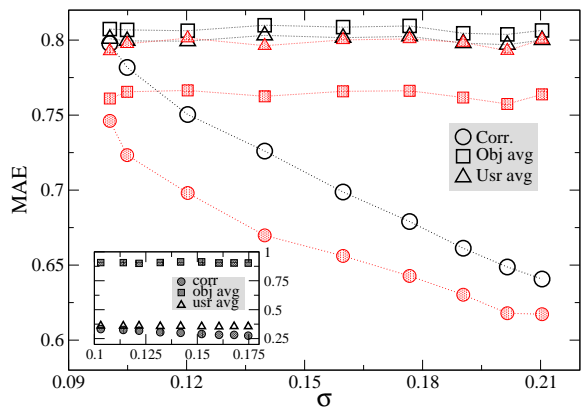}}
\caption{{\sl Main graph}: Mean prediction error of
  three recommending methods (user average, item average,
  correlation-based), as a function of the standard
  deviation $\sigma$ of user-user correlations. 
The dimension of the data set, generated artificially as in section~\ref{simu},
is $N=250$, $M=500$.
  The correlation matrix is distributed uniformly,
  with mean $\mu=0$ (empty symbols) and $\mu=0.1$ (filled
  symbols). The lines are only meant to guide the eye.
{\sl Inset}: Same as the main graph, with $\mu=0.1$, but 
  here the votes follow a bimodal distribution. As a consequence, the
  user average $\avg{v_i}$ works better than the item average
  $m(\alpha)$.
}
\label{fig:art1}
\end{figure}

In fig.~\ref{fig:art1} we plot, for two different values of $\mu$, 
the MAE of the predictions as a function of $\sigma$. In the main
graph, open and filled symbols are simulations performed with
$\mu=0$ and $\mu=0.1$ respectively. As expected, the
performance of user
averages $\avg{v_i}$, represented by triangles, does not depend on the parameters
of the distribution of correlations --as long as the distribution of
votes is unimodal. On the contrary the object average $m(\alpha)$, represented by squares,
obviously improves when $\mu$ increases, but it is still independent of $\sigma$.
The correlation method also works better for $\mu=0.1$ (filled
circles) than for $\mu=0$ (empty circles). Its MAE diminishes as well
for increasing $\sigma$. 
When $\sigma$ goes to zero, in fact, all pair correlations are equal $C_{i j}=c
\quad \forall i,j$ and the correlation-weighted sums in eq.~\req{pred}
become proportional to $m(\alpha)$. This appears clearly
in fig.~\ref{fig:art1}, where circles and squares tend to overlap for
$\sigma\to 0$. A similar situation occurs in the MovieLens database,
where both $\mu$ and $\sigma$ are very small. It is not surprising
that, in the sparse region of fig.~\ref{fig1}, the mean becomes an even
better predictor.

Votes in the Jester database can be better predicted by user averages
$\avg{v_i}$ than by item averages $m(\alpha)$. The reason for this is
that users are grouped, as shown in fig.~\ref{jeseig}.
The distribution of
Jester votes is, as appears in fig.~\req{fig0}, slightly bimodal. We
have generated a data set with this additional feature and plotted the result in the inset of
fig.~\ref{fig:art1}. We obtain a behaviour that is similar to
that of fig.~\ref{fig2}, confirming our hypothesis.

\section{Conclusions}
\label{con}
We have introduced a new method for recommending, based on the
spectrum of the normalized Laplacian of a weighted graph in the user
space. This has been tested on the MovieLens and Jester databases,
together with a refined correlation-based method. 

The experiments have
been made on raw data, without altering the statistical properties of
the original voting matrix. In particular, no restriction has been
imposed on the minimal number of votes expressed by users or received
by items. For Movielens, the opinions have been ordered according to
their timestamp.

Our spectral method proves to be the most effective in all cases
considered. The predictive power of recommender systems 
is stronger in the Jester than in the MovieLens case, where simple
averages are able to detect most information contained in the data.
We argue that this phenomenon is due, at least in
part, to a different distribution of user-user correlations. When the latter
is broader, in fact, sophisticated methods are much more
rewarding. A distinction between unimodal and bimodal distributions of
votes has been made to determine the best way to take simple averages.

In conclusion, our findings can be used to determine whether or not it
is worth to develop complex methods for recommending in  specific
contexts.

\appendix
\section{How to draw a valid correlation matrix}
\label{app1}
Here we explain the procedure, inspired by ref.~\cite{JaRe99}, to
create the correlation matrix $\bar{C}$ used to draw a
multivariate Gaussian distribution of votes.
The pair correlation of the population of users
must follow a given distribution 
with the desired mean $\mu$ and variance $\sigma$.
Moreover, the matrix we are looking for must be positive semi-definite,
i.e. $\lambda_{i} \ge 0 \quad \forall i$, where the $\lambda_{i}$'s are the
eigenvalues of $\bar{C}$. Let us construct it step by step.

I) We create a square matrix $A(N \times N)$, with elements
$A_{ij}$ drawn from a given distribution (uniform in our simulations) of
mean $\mu$ and variance $\sigma$. $A$ will not be symmetric in
general. II) We
apply the transformation $\bar{A} = U + U^{T}$,
where $U$ is the upper triangular matrix of $A$ and $U^{T}$ its transposed. $\bar{A}$
is now symmetric, but not positive semi-definite.
III) We calculate the right eigensystem $E$ of the real symmetric matrix $\bar{A}$
and its associated set of eigenvalues $\{\lambda_{i}\}$. Hence
$\bar{A} \cdot C = \Lambda \cdot E$, with $\Lambda =
diag(\lambda_{i})$. Some eigenvalues can be netative. IV) Let us define
$\lambda^{'}_{i}=\lambda_{i} \quad
\forall \lambda_{i}>0$, $\lambda^{'}_{i}=0$ otherwise.
The diagonal matrix $\Lambda'$ has then semi-positive elements $\lambda^{'}_{i}$. 
V) Given the scaling matrix $T_{i} = \left[ \sum_{m} s_{im}^{2}
  \lambda^{'}_{m} \right]^{-1}$ and the matrices
$B'= E \sqrt{\Lambda'}$ and $B = \sqrt{T} E \sqrt{\Lambda'}$,
the matrix $\bar{C} = BB^{T}$ is positive-semidefinite and
has unit diagonal elements by contstruction, but not the desired mean and
variance we imposed on the original matrix $A$. VI) By adding a
constant value to the elements of $\bar{C}$, we adjust
its mean to $\mu$ and its standard deviation to $\sigma$. 
We rename the matrix thus obtained $\bar{A}$ and iterate the algorithm,
from step III, till the good values of $\mu$ and $\sigma$
are obtained for $\bar{C}$ with the desired precision.

\begin{acks}
The authors would like to aknowledge the inspiring guide of
Yi-Cheng Zhang, as well as the useful discussions they had with Lionel Moret and
Hassan Masum. This work has been supported by the
Swiss National Science Foundation under grant number 205120-113842.
\end{acks}

\bibliographystyle{acmtrans}
\bibliography{mvjest}

\end{document}